\documentclass{JHEP3}
\usepackage{epsfig}
\usepackage{graphicx}%
\def\be{\begin{equation}}
\def\ee{\end{equation}}
\def\bea{\begin{eqnarray}}
\def\eea{\end{eqnarray}}

\def\lesssim{\mathrel{\hbox{\rlap{\hbox{\lower4pt\hbox{$\sim$}}}\hbox{$<$}}}}
\def\gtrsim{\mathrel{\hbox{\rlap{\hbox{\lower4pt\hbox{$\sim$}}}\hbox{$>$}}}}

\title{Prospects of Inflation\footnote{Extended version of the talk at the Nobel
Symposium ``Cosmology \& String Theory," August 2003}}

\author{Andrei Linde\\
    Department of Physics, Stanford University, Stanford, CA 94305,
USA\\
    E-mail: \email{alinde@stanford.edu}}

\received{\today} \preprint{SU-ITP-04/04\\
\hepth{0402051}\\ February 6, 2004}

\abstract{I will discuss the development of inflationary theory
and its present status, including recent progress in describing de
Sitter space and inflationary universe in string theory.}

\begin{document}

\section{Introduction}

After more than 20 years of its existence, inflationary theory
gradually becomes  the standard cosmological paradigm. However, we
still do not know which of the many versions of inflationary
cosmology will be favored by the future observational data.
Moreover, it may be quite nontrivial to obtain a natural
realization of inflationary theory in the context of the ever
changing theory of all fundamental interactions.

Development of inflationary cosmology occurs in many different
ways. The original tendency was to concentrate on the basic
principles of inflationary theory, such as the problem of initial
conditions, the possibility of eternal inflation, etc. Recent
progress in observational cosmology shifted attention towards
experimental verification of various inflationary theories. For
example, one can parametrize inflationary theories by several
slow-roll parameters and find the observational constraints on
these parameters. Another approach is to look  for inflationary
models which can work in the context of the simplest
phenomenological theories of elementary particles. A more
ambitious trend  is to implement inflationary cosmology in string
theory. Because of this multitude of goals and methods, different
people sometimes have rather different ideas about inflationary
theory and its future prospects. In this paper we will try to
analyze the situation.

\section{Brief history of inflation}

The first model of inflationary type was proposed by Alexei
Starobinsky \cite{Star}. It was based on investigation of
conformal anomaly in quantum gravity. This model  did not suffer
from the graceful exit problem, but it was rather complicated and
did not aim on solving homogeneity, horizon and monopole problems.

A much simpler inflationary model with a very clear physical
motivation was proposed by Alan Guth \cite{Guth}.  His model,
which is now called ``old inflation,'' was based on the theory of
supercooling during the cosmological phase transitions
\cite{Kirzhnits}. Even though this scenario did not work,  it
played a profound role in the development of inflationary
cosmology since it contained a very clear explanation how
inflation may solve the major cosmological problems.

According to this scenario,  inflation is as   exponential
expansion of the universe in a supercooled false vacuum state.
False vacuum is a metastable state without any fields or particles
but with large energy density. Imagine a universe filled with such
``heavy nothing.'' When the universe expands, empty space remains
empty, so its energy density does not change. The universe with a
constant energy density expands exponentially, thus we have
inflation in the false vacuum. Then the false vacuum decay, the
bubbles of the new phase collide, and our universe becomes hot.

Unfortunately, if the bubbles of the new phase are formed near
each other, their collisions make the universe extremely
inhomogeneous. If they are formed far away from each other, each
of them represents a separate open universe with a vanishingly
small $\Omega$. Both options are unacceptable  \cite{Guth}.

This problem was resolved with the invention of the new
inflationary theory \cite{New}. In this theory, inflation may
begin either in the false vacuum,  or in an unstable state at the
top of the effective potential. Then the inflaton field $\phi$
slowly rolls down to the minimum of its effective potential.  The
motion of the field away from the false vacuum is of crucial
importance: density perturbations produced during the slow-roll
inflation are inversely proportional to $\dot \phi$
\cite{Mukh,Hawk,Mukh2}. Thus the key difference between the new
inflationary scenario and the old one is that the useful part of
inflation in the new scenario, which is responsible for the
homogeneity of our universe, does {\it not} occur in the false
vacuum state, where $\dot\phi =0$.

This scenario was so popular in the beginning of the 80's that
even now most textbooks on astrophysics describe inflation as an
exponential expansion during high temperature phase transitions in
grand unified theories. Unfortunately, the new inflation scenario
was plagued by its own problems. It works only if the effective
potential of the field $\phi$ has a very a flat plateau near $\phi
= 0$, which is somewhat artificial. In most versions of this
scenario the inflaton field has an extremely small coupling
constant, so it could not be in a thermal equilibrium with other
matter fields. The theory of cosmological phase transitions, which
was the basis for old and new inflation, did not work in such a
situation. Moreover, thermal equilibrium requires many particles
interacting with each other. This means that new inflation could
explain why our universe was so large only if it was very large
and contained many particles from the very beginning. Finally,
inflation in this theory begins very late. During the preceding
epoch the universe can easily collapse or become so inhomogeneous
that inflation may never happen \cite{book}. Because of all of
these difficulties, no realistic versions of the new inflationary
universe scenario have been proposed so far.

Old and new inflation represented  a substantial but  incomplete
modification of the big bang theory. It was still assumed that the
universe was in a state of thermal equilibrium from the very
beginning, that it was relatively homogeneous and large enough to
survive until the beginning of inflation, and that the stage of
inflation was just an intermediate stage of the evolution of the
universe. In the beginning of the 80's these assumptions seemed
most natural and practically unavoidable. On the basis of all
available observations (CMB, abundance of light elements)
everybody believed that the universe was created in a hot big
bang. That is why it was so difficult to overcome a certain
psychological barrier and abandon all of these assumptions. This
was done with the invention of the chaotic inflation scenario
\cite{Chaot}. This scenario resolved all problems of old and new
inflation. According to this scenario, inflation may occur even in
the theories with simplest potentials such as $V(\phi) \sim
\phi^n$. Inflation may begin even if there was no thermal
equilibrium in the early universe, and it may start even at the
Planckian density, in which case the problem of initial conditions
for inflation can be easily resolved \cite{book}.

\section{Chaotic Inflation}

Consider  the simplest model of a scalar field $\phi$ with a mass
$m$ and with the potential energy density $V(\phi)  = {m^2\over 2}
\phi^2$. Since this function has a minimum at $\phi = 0$,  one may
expect that the scalar field $\phi$ should oscillate near this
minimum. This is indeed the case if the universe does not expand,
in which case equation of motion for the scalar field  coincides
with equation for harmonic oscillator, $\ddot\phi = -m^2\phi$.

However, because of the expansion of the universe with Hubble
constant $H = \dot a/a $, an additional  term $3H\dot\phi$ appears
in the harmonic oscillator equation:
\begin{equation}\label{1}
 \ddot\phi + 3H\dot\phi = -m^2\phi \ .
\end{equation}
The term $3H\dot\phi$ can be interpreted as a friction term. The
Einstein equation for a homogeneous universe containing scalar
field $\phi$ looks as follows:
\begin{equation}\label{2}
H^2 +{k\over a^2} ={1\over 6}\, \left(\dot \phi ^2+m^2 \phi^2)
\right) \ .
\end{equation}
Here $k = -1, 0, 1$ for an open, flat or closed universe
respectively. We work in units $M_p^{-2} = 8\pi G = 1$.

If   the scalar field $\phi$  initially was large,   the Hubble
parameter $H$ was large too, according to the second equation.
This means that the friction term $3H\dot\phi$ was very large, and
therefore    the scalar field was moving   very slowly, as a ball
in a viscous liquid. Therefore at this stage the energy density of
the scalar field, unlike the  density of ordinary matter, remained
almost constant, and expansion of the universe continued with a
much greater speed than in the old cosmological theory. Due to the
rapid growth of the scale of the universe and a slow motion of the
field $\phi$, soon after the beginning of this regime one has
$\ddot\phi \ll 3H\dot\phi$, $H^2 \gg {k\over a^2}$, $ \dot \phi
^2\ll m^2\phi^2$, so  the system of equations can be simplified:
\begin{equation}\label{E04}
H= {\dot a \over a}   ={ m\phi\over \sqrt 6}\ , ~~~~~~  \dot\phi =
-m\  \sqrt{2\over 3}     .
\end{equation}
The first equation shows that if the field $\phi$ changes slowly,
the size of the universe in this regime grows approximately as
$e^{Ht}$, where $H = {m\phi\over\sqrt 6}$. This is the stage of
inflation, which ends when the field $\phi$ becomes much smaller
than $M_p=1$. Solution of these equations shows that after a long
stage of inflation  the universe initially filled with the field
$\phi = \phi_0 \gg 1$  grows  exponentially \cite{book}, $ a= a_0
\ e^{\phi_0^2/4}$.

This is as simple as it could be. Inflation does not require
supercooling and tunnelling from the false  vacuum \cite{Guth}, or
rolling from an artificially flat top of the effective potential
\cite{New}. It appears in the theories that can be as simple as a
theory of a harmonic oscillator \cite{Chaot}. Only after I
realized it, I started to believe that inflation is not a trick
necessary to fix  problems of the old big bang theory, but a
generic cosmological regime.

In  realistic versions of inflationary theory the  duration of
inflation could be as short as $10^{-35}$ seconds. When inflation
ends, the scalar field $\phi$ begins to   oscillate near the
minimum of $V(\phi)$. As any rapidly oscillating classical field,
it looses its energy by creating pairs of elementary particles.
These particles interact with each other and come to a state of
thermal equilibrium with some temperature $T$ \cite{DL,KLS,tach}.
From this time on, the universe can be described by the usual big
bang theory.

The main difference between inflationary theory and the old
cosmology becomes clear when one calculates the size of a typical
inflationary domain at the end of inflation. Investigation of this
question    shows that even if  the initial size of   inflationary
universe  was as small as the Planck size $l_P \sim 10^{-33}$ cm,
after $10^{-35}$ seconds of inflation   the universe acquires a
huge size of   $l \sim 10^{10^{12}}$ cm! This number is
model-dependent, but in all realistic models the  size of the
universe after inflation appears to be many orders of magnitude
greater than the size of the part of the universe which we can see
now, $l \sim 10^{28}$ cm. This immediately solves most of the
problems of the old cosmological theory \cite{Chaot,book}.

Our universe is almost exactly homogeneous on  large scale because
all inhomogeneities were exponentially stretched during inflation.
The density of  primordial monopoles  and other undesirable
``defects'' becomes exponentially diluted by inflation.   The
universe   becomes enormously large. Even if it was a closed
universe of a size
 $\sim 10^{-33}$ cm, after inflation the distance between its ``South'' and
``North'' poles becomes many orders of magnitude greater than
$10^{28}$ cm. We see only a tiny part of the huge cosmic balloon.
That is why nobody  has ever seen how parallel lines cross. That
is why the universe looks so flat.

If our universe initially consisted of many domains with
chaotically distributed scalar field  $\phi$ (or if one considers
different universes with different values of the field), then
domains in which the scalar field was too small never inflated.
The main contribution to the total volume of the universe will be
given by those domains which originally contained large scalar
field $\phi$. Inflation of such domains creates huge homogeneous
islands out of initial chaos. (That is why I called this scenario
``chaotic inflation.'') Each  homogeneous domain in this scenario
is much greater than the size of the observable part of the
universe.

The first models of chaotic inflation were based on the theories
with polynomial potentials, such as $V(\phi) = \pm {m^2\over 2}
\phi^2 +{\lambda\over 4} \phi^4$. But the main idea of this
scenario is quite generic. One should consider any particular
potential $V(\phi)$, polynomial or not, with or without
spontaneous symmetry breaking, and study all possible initial
conditions without assuming that the universe was in a state of
thermal equilibrium, and that the field $\phi$ was in the minimum
of its effective potential from the very beginning.

This scenario strongly deviated from the standard lore of the hot
big bang theory and was psychologically difficult to accept.
Therefore during the first few years after invention of chaotic
inflation many authors claimed that the idea of chaotic initial
conditions is unnatural, and made attempts to realize the new
inflation scenario based on the theory of high-temperature phase
transitions, despite numerous problems associated with it. Some
authors believed that the theory must satisfy so-called `thermal
constraints' which were necessary to ensure that the minimum of
the effective potential at large $T$ should be at $\phi=0$
\cite{OvrStein}, even though the scalar field in the models  they
considered was not in a state of thermal equilibrium with other
particles. It took several years until it finally became clear
that the idea of chaotic initial conditions is most general, and
it is much easier to construct a consistent cosmological theory
without making unnecessary assumptions about thermal equilibrium
and high temperature phase transitions in the early universe.

A short note of terminology:  Chaotic inflation occurs in all
models with sufficiently flat potentials, including the potentials
originally used in new inflation \cite{Linde:cd}.  The versions of
chaotic inflation with such potentials for simplicity are often
called `new inflation', even though inflation begins there not as
in the original new inflation scenario (i.e. not due to the phase
transitions with supercooling), but as in the chaotic inflation
scenario.

\section{Hybrid inflation}
In the previous section we considered the simplest chaotic
inflation theory based on  the theory of a single scalar field
$\phi$. The models of chaotic inflation based on the theory of two
scalar fields may have some qualitatively new features. One of the
most interesting models of this kind is the hybrid inflation
scenario \cite{Hybrid}.  The simplest version of this scenario is
based on chaotic inflation in the theory of two scalar fields with
the effective potential
\begin{equation}\label{hybrid}
V(\sigma,\phi) =  {1\over 4\lambda}(M^2-\lambda\sigma^2)^2 +
{m^2\over 2}\phi^2 + {g^2\over 2}\phi^2\sigma^2\ .
\end{equation}
The effective mass squared of the field $\sigma$ is equal to $-M^2
+ g^2\phi^2$.  Therefore for $\phi > \phi_c = M/g$ the only
minimum of the effective potential $V(\sigma,\phi)$ is at $\sigma
= 0$. The curvature of the effective potential in the
$\sigma$-direction is much greater than in the $\phi$-direction.
Thus  at the first stages of expansion of the universe the field
$\sigma$ rolled down to $\sigma = 0$, whereas the field $\phi$
could remain large for a much longer time.

At the moment when the inflaton field $\phi$ becomes smaller than
$\phi_c = M/g$,  the phase transition with the symmetry breaking
occurs. If $m^2 \phi_c^2 = m^2M^2/g^2 \ll M^4/\lambda$, the Hubble
constant at the time of the phase transition is given by $H^2 =
{M^4 \over 12 \lambda}$ (in units $M_ = 1$). If one assumes that
$M^2 \gg {\lambda m^2\over g^2}$ and that $m^2 \ll H^2$, then the
universe at $\phi > \phi_c$ undergoes a stage of inflation, which
abruptly ends at $\phi = \phi_c$.

One of the advantages of this scenario is the possibility to
obtain small density  perturbations even if coupling constants are
large, $\lambda, g  = O(1)$. This scenario works   if the
effective potential has a relatively flat $\phi$-direction. But
flat directions often appear in supersymmetric theories. This
makes hybrid inflation an attractive playground for those who
wants to achieve inflation in supergravity and string theory. We
will return to this question later.

\section{Quantum fluctuations and density perturbations
\label{Perturb}}

The vacuum structure in the  exponentially expanding universe is
much more complicated than in ordinary Minkowski space.
 The wavelengths of all vacuum
fluctuations of the scalar field $\phi$ grow exponentially during
inflation. When the wavelength of any particular fluctuation
becomes greater than $H^{-1}$, this fluctuation stops oscillating,
and its amplitude freezes at some nonzero value $\delta\phi (x)$
because of the large friction term $3H\dot{\phi}$ in the equation
of motion of the field $\phi$\@. The amplitude of this fluctuation
then remains almost unchanged for a very long time, whereas its
wavelength grows exponentially. Therefore, the appearance of such
a frozen fluctuation is equivalent to the appearance of a
classical field $\delta\phi (x)$ that does not vanish after
averaging over macroscopic intervals of space and time.

Because the vacuum contains fluctuations of all wavelengths,
inflation leads to the creation of more and more new perturbations
of the classical field with wavelengths greater than $H^{-1}$\@.
The average amplitude of such perturbations generated during a
typical time interval $H^{-1}$ is given by
\cite{Vilenkin:wt,Linde:uu}
\begin{equation}\label{E23}
|\delta\phi(x)| \approx \frac{H}{2\pi}\ .
\end{equation}

These fluctuations lead to density perturbations that later
produce galaxies. The theory of this effect  is very complicated
\cite{Mukh,Hawk}, and it was fully understood only in the second
part of the 80's \cite{Mukh2}. The main idea can be described as
follows:

Fluctuations of the field $\phi$ lead to a local delay of the time
of the end of inflation,  $\delta t = {\delta\phi\over \dot\phi}
\sim {H\over 2\pi \dot \phi}$. Once the usual post-inflationary
stage begins, the density of the universe starts to decrease as
$\rho = 3 H^2$, where $H \sim t^{-1}$. Therefore a local delay of
expansion leads to a local density increase $\delta_H$ such that
$\delta_H \sim \delta\rho/\rho \sim  {\delta t/t}$. Combining
these estimates together yields the famous result
\cite{Mukh,Hawk,Mukh2}
\begin{equation}\label{E24}
\delta_H \sim \frac{\delta\rho}{\rho} \sim {H^2\over 2\pi\dot\phi}
\ .
\end{equation}
This derivation is oversimplified; it does not tell, in
particular, whether $H$ should be calculated during inflation or
after it. This issue was not very important for new inflation
where $H$ was nearly constant, but it is of crucial importance for
chaotic inflation.

The result of a more detailed investigation \cite{Mukh2} shows
that $H$ and $\dot\phi$ should be  calculated during inflation, at
different times for perturbations with different momenta $k$. For
each of these perturbations the value of $H$ should be taken at
the time when the wavelength of the perturbation  becomes of the
order of $H^{-1}$. However, the field $\phi$ during inflation
changes very slowly, so the quantity ${H^2\over 2\pi\dot\phi}$
remains almost constant over exponentially large range of
wavelengths. This means that the spectrum of perturbations of
metric is flat.

A detailed calculation in our simplest chaotic inflation model the
amplitude of perturbations gives
\begin{equation}\label{E26}
\delta_H \sim   {m \phi^2\over 5\pi \sqrt 6} \ .
\end{equation}
The perturbations on scale of the horizon were produced at
$\phi_H\sim 15$ \cite{book}. This, together  with COBE
normalization $\delta_H \sim 2 \times 10^{-5}$  gives $m \sim
3\times 10^{-6}$, in Planck units, which is approximately
equivalent to $7 \times 10^{12}$ GeV. Exact numbers depend on
$\phi_H$, which in its turn depends slightly on the subsequent
thermal history of the universe.

The magnitude of density perturbations $\delta_H$ in our model
depends on the scale $l$ only logarithmically. Since the
observations provide us with an information about a rather limited
range of $l$, it is possible to parametrize the scale dependence
of density perturbations by a simple power law, $\delta_H \sim
l^{(1-n)/2}$. An exactly flat spectrum would correspond to $n =
1$.

Flatness of the spectrum of density perturbations,  together with
flatness of the universe  ($\Omega = 1$),  constitute the two most
robust predictions of inflationary cosmology. It is possible to
construct models where $\delta_H$ changes in a very peculiar way,
and it is also possible to construct theories where $\Omega \not =
1$, but it is difficult to do so.

\section{Eternal inflation}

A significant step in the development of inflationary theory was
the discovery of the process of self-reproduction of inflationary
universe. This process was known to exist in old inflationary
theory \cite{Guth} and in the new one \cite{Vilenkin:xq}, but its
significance was fully realized only after the discovery of the
regime of eternal inflation in the simplest versions of the
chaotic inflation scenario \cite{Eternal,LLM}. It appears that in
many models large quantum fluctuations produced during inflation
which may locally increase the value of the energy density in some
parts of the universe. These regions expand at a greater rate than
their parent domains, and quantum fluctuations inside them lead to
production of new inflationary domains which expand even faster.
This  leads to an eternal process of self-reproduction of the
universe.

To understand the mechanism of self-reproduction one should
remember that the processes separated by distances $l$ greater
than $H^{-1}$ proceed independently of one another. This is so
because during exponential expansion the distance between any two
objects separated by more than $H^{-1}$ is growing with a speed
exceeding the speed of light. As a result, an observer in the
inflationary universe can see only the processes occurring inside
the horizon of the radius  $H^{-1}$. An important consequence of
this general result is that the process of inflation in any
spatial domain of radius $H^{-1}$ occurs independently of any
events outside it. In this sense any inflationary domain of
initial radius exceeding $H^{-1}$ can be considered as a separate
mini-universe.

To investigate the behavior of such a mini-universe, with an
account taken of quantum fluctuations, let us consider an
inflationary domain of initial radius $H^{-1}$ containing
sufficiently homogeneous field with initial value $\phi \gg M_p$.
Equation (\ref{E04}) implies that during a typical time interval
$\Delta t=H^{-1}$ the field inside this domain will be reduced by
$\Delta\phi = \frac{2}{\phi}$. By comparison this expression with
$|\delta\phi(x)| \approx \frac{H}{2\pi} =  {m\phi\over 2\pi\sqrt
6}$ one can easily see that if $\phi$ is much less than $\phi^*
\sim {5\over  \sqrt{ m}} $,
 then the decrease of the field $\phi$
due to its classical motion is much greater than the average
amplitude of the quantum fluctuations $\delta\phi$ generated
during the same time. But for   $\phi \gg \phi^*$ one has
$\delta\phi (x) \gg \Delta\phi$. Because the typical wavelength of
the fluctuations $\delta\phi (x)$ generated during the time is
$H^{-1}$, the whole domain after $\Delta t = H^{-1}$ effectively
becomes divided into $e^3 \sim 20$ separate domains
(mini-universes) of radius $H^{-1}$, each containing almost
homogeneous field $\phi - \Delta\phi+\delta\phi$.   In almost a
half of these domains the field $\phi$ grows by
$|\delta\phi(x)|-\Delta\phi \approx |\delta\phi (x)| = H/2\pi$,
rather than decreases. This means that the total volume of the
universe containing {\it growing} field $\phi$ increases 10 times.
During the next time interval $\Delta t = H^{-1}$ the situates
repeats. Thus, after the two time  intervals $H^{-1}$ the total
volume of the universe containing the growing scalar field
increases 100 times, etc. The universe enters eternal process of
self-reproduction.

Realistic models of elementary particles involve many kinds of
scalar fields.  For example, in the unified theories of weak,
strong and electromagnetic interactions, at least two other scalar
fields exist. The potential energy of these scalar fields may have
several different minima. This means that the same theory may have
different vacuum states, corresponding to different types of
symmetry breaking between fundamental interactions, and, as a
result, to different laws of low-energy physics.

During the process of eternal inflation in the simplest versions
of the chaotic inflation scenario (but not in the new inflation),
some parts of the universe spend indefinitely long time at the
nearly Planckian density, expanding with the Hubble constant
$H=O(1)$, in Planck mass units. In this regime, all scalar fields
persistently experience quantum jumps of the magnitude comparable
to the Planck mass. This forces the fields to browse between all
possible vacuum states. As a result, the universe  becomes divided
into infinitely many exponentially large domains that have
different laws of low-energy physics \cite{Eternal,LLM}. This
result may have especially interesting implications in the context
of string theory, which allows exponentially large number of
different vacuum states \cite{book,landscape,Douglas}, see
 Sect. \ref{land}.

\section{Inflation and observations}

Inflation is not just an interesting theory that can resolve many
difficult problems of the standard  Big Bang cosmology. This
theory made several   predictions which can be tested by
cosmological observations. Here are the most important
predictions:

1) The universe must be flat. In most models $\Omega_{total} = 1
\pm 10^{-4}$.

2) Perturbations of metric produced during inflation are
adiabatic.

3) Inflationary perturbations have flat spectrum.  In most
inflationary models the spectral index $n = 1 \pm 0.2$ ($n=1$
means totally flat.)

4) These perturbations are gaussian.

5) Perturbations of metric could be scalar, vector or tensor.
Inflation mostly produces scalar  perturbations, but it also
produces tensor perturbations with nearly flat spectrum, and it
does {\it not} produce vector perturbations. There are certain
relations between the properties of  scalar and tensor
perturbations produced by inflation.

6) Inflationary perturbations produce specific peaks in the
spectrum of CMB radiation. (For a simple pedagogical
interpretation of this effect see  e.g. \cite{Dodelson:2003ip}; a
detailed theoretical description can be found in
\cite{Mukhanov:2003xr}.)

It is possible to violate each of these predictions if one makes
this theory sufficiently complicated. For example, it is possible
to produce vector perturbations of metric in the models where
cosmic strings are produced at the end of inflation, which is the
case in some versions of hybrid inflation. It is possible to have
an open or closed inflationary universe, or even a small periodic
inflationary universe, it is possible to have models with
nongaussian isocurvature fluctuations with a non-flat spectrum.
However, it is very difficult to do so, and most of the
inflationary models satisfy the simple rules given above.

It is not easy to test all of these predictions. The major
breakthrough in this direction was achieved  due to the recent
measurements of the CMB anisotropy. The latest results based on
the WMAP experiment, in combination with the Sloan Digital Sky
Survey, are consistent with predictions of the simplest
inflationary models with adiabatic gaussian perturbations, with
$\Omega = 1.01 \pm 0.02$, and $n = 0.98 \pm
0.03$~\cite{WMAP,Tegmark}.

There are still some question marks to be examined, such as the
unexpectedly small anisotropy of CMB at  large angles \cite{WMAP}.
It is not quite clear whether we  deal with a real anomaly here or
with a manifestation of cosmic variance \cite{Efstathiou:2003wr},
but in any case, it is quite significant that all proposed
resolutions of this problem are based on inflationary cosmology,
see e.g. \cite{Contaldi}.

\section{Alternatives to inflation?}\label{alt}

Inflationary scenario is very versatile, and now, after 20 years
of persistent attempts of many physicists to propose an
alternative to inflation, we still do not know any other  way to
construct a consistent cosmological theory. Indeed, in order to
compete with inflation a new theory should make similar
predictions and should offer an alternative solution to many
difficult cosmological problems. Let us look at these problems
before starting a discussion.

1) Homogeneity problem. Before even starting investigation of
density perturbations and structure  formation, one should explain
why the universe is nearly homogeneous on the horizon scale.

2) Isotropy problem. We need to understand why all directions in
the universe are similar to each  other, why there is no overall
rotation of the universe. etc.

3) Horizon problem. This one is closely related to the homogeneity
problem. If different parts of  the universe have not been in a
causal contact when the universe was born, why do they look so
similar?

4) Flatness problem. Why $\Omega \approx 1$? Why parallel lines do
not intersect?

5) Total entropy problem. The total entropy of the observable part
of the universe is  greater than $10^{87}$. Where did this huge
number come from? Note that the lifetime of a closed universe
filled with hot gas with total entropy $S$  is $S^{2/3}\times
10^{-43}$ seconds \cite{book}. Thus $S$ must be huge. Why?

6) Total mass problem. The total mass of the observable part of
the universe has mass  $\sim 10^{60} M_p$.  Note also that the
lifetime of a closed universe filled with nonrelativistic
particles of total mass $M$ is ${M\over M_P} \times 10^{-43}$
seconds. Thus $M$ must be huge. But why?

7) Structure formation problem. If we manage to explain the
homogeneity of the universe, how can  we explain the origin of
inhomogeneities required for the large scale structure formation?

8) Monopole problem, gravitino problem, etc.

This list is very long. That is why it was not easy to propose any
alternative to inflation even  before we learned that $\Omega
\approx 1$, $n\approx 1$, and that the perturbations responsible
for galaxy formation are mostly adiabatic, in agreement with the
predictions of the simplest inflationary models.

Despite this difficulty, there was always a tendency to announce
that we have  eventually found a good alternative to inflation.
This was the ideology of the models of structure formation due to
topological defects, or textures, which were often described as
competitors to inflation, see e.g. \cite{SperTur}. However, it was
clear from the very beginning that these theories at best could
solve only one problem (structure formation) out of 8 problems
mentioned above. Therefore the true question was not whether one
can replace inflation by the theory of cosmic strings/textures,
but whether inflation with cosmic strings/textures is better than
inflation without cosmic strings/textures. Recent observational
data favor the simplest version of inflationary theory, without
topological defects, or with an extremely small (few percent)
admixture of the effects due to cosmic strings.

A similar situation emerged recently with the introduction of the
ekpyrotic  scenario  \cite{KOST}. In the original version of this
theory it was claimed that this scenario can solve all
cosmological problems without using the stage of inflation, i.e.
without a prolonged stage of an accelerated expansion of the
universe, which was called in \cite{KOST} ``superluminal
expansion.'' However, this original idea did not work
\cite{KKL,KKLTS}, and the idea to avoid ``superluminal expansion''
was abandoned by the authors of \cite{KOST}. A more recent version
of this scenario, the cyclic scenario \cite{cyclic}, uses an
infinite number of periods of ``superluminal expansion'' (i.e.
inflation) in order to solve the major cosmological problems. The
main difference between the usual inflation and the cyclic
inflation, just as in the case of topological defects and
textures, is the mechanism of generation of density perturbations.
However, since the theory of density perturbations in cyclic
inflation requires a solution of the cosmological singularity
problem \cite{Liu:2002ft,Horowitz:2002mw}, it is difficult to say
anything definite about it.

Thus at the moment it is hard to see any real alternative to
inflationary cosmology; instead of a competition between inflation
and other ideas, we witness a competition between many different
models of inflationary theory.

This competition goes in several different directions. First of
all, we must try to implement inflation in realistic theories of
fundamental interactions. But what do we mean by `realistic?'
In the absence of a direct confirmation of M/string theory and
supergravity by high energy physics experiments (which may change
when we start receiving data from the LHC), the definition of what
is realistic becomes increasingly dependent on cosmology and the
results of the cosmological observations. In particular, one may
argue that those versions of the theory of all fundamental
interactions that cannot describe inflation and the present stage
of acceleration of the universe are disfavored by observations.

On the other hand, not every theory which can lead to inflation
does it in an equally good way. Many inflationary models have been
already ruled out be observations. This happened long ago with
such models as extended inflation \cite{extended} and the simplest
versions of ``natural inflation'' \cite{natural}. Recent data from
WMAP and SDSS almost ruled out a particular version of chaotic
inflation with $V(\phi) \sim \phi^4$ \cite{WMAP,Tegmark}.

However, observations test only the last stages of inflation. In
particular, they  do not say anything about the properties of the
inflaton potential at $V(\phi) \gtrsim 10^{-10} M_p^4$. Thus there
may exist many different models which describe all observational
data equally well. In order to compare such models, one should not
only compare their predictions with the results of the
cosmological observations, but also carefully examine whether they
really solve the main cosmological problems.

One of the most important issues here is the maximal value of
energy density during inflation. For example, the simplest chaotic
inflation scenario may begin in the universe of the largest
possible energy density, of a smallest possible size (Planck
length), with the smallest possible mass $M \sim M_p$ and with the
smallest possible entropy $S = O(1)$. This provides a true
solution to the flatness, horizon, homogeneity, mass and entropy
problems. Meanwhile, in the new inflation scenario, inflation
occurs on the mass scale 3 orders of magnitude below $M_p$, when
the total size of the universe must be very large. As a result,
the initial mass of the universe at the beginning of new inflation
must be  greater than $10^6 M_p$, and its total entropy must be
greater than $10^9$. In other words, in order to explain why the
entropy of the universe is greater than $10^{87}$ one should
assume that it was extremely large from the very beginning. This
does not look like a real solution of the entropy problem. A
similar problem exists in many of the models recently described in
\cite{Lyth:2003kp}. Finally, in cyclic inflation, the process of
exponential expansion of the universe occurs only if the total
mass of the universe is greater than its present mass $M \sim
10^{60} M_p$ and its total entropy is greater than $10^{87}$. This
scenario does not solve the flatness, mass and entropy problems.

One can argue  \cite{LLM} that eternal inflation may alleviate
some of the problems of the low-scale inflation. Note, however,
that eternal inflation, which naturally occurs in the simplest
versions of chaotic inflation and in new inflation, does not exist
in many popular versions of hybrid inflation. Of course,  models
of low-scale non-eternal inflation are still much better than the
models with no inflation at all, but I do not think that we should
settle for the second-best. It would be much better to have a
stage of eternal chaotic inflation at nearly Planckian density,
which, if needed, may be followed by a stage of low-scale
inflation.

Keeping eternal inflation at high density as our ultimate goal,
let discuss a possibility to obtain inflation in supergravity and
string theory.

\section{Shift symmetry and chaotic inflation in supergravity}

Most of the existing inflationary models are based on the idea of
chaotic initial conditions, which is the trademark of the chaotic
inflation scenario. In the simplest versions of chaotic inflation
scenario with the potentials $V \sim \phi^n$, the process of
inflation occurs at $\phi>1$, in Planck units. Meanwhile, there
are many other models where inflation may occur at $\phi \ll 1$.

There are several reasons why this difference may be important.
First of all, some authors argue that  the generic expression for
the effective potential can be cast in the form
\begin{equation}\label{LythRiotto}
V(\phi) = V_0 +\alpha \phi+ {m^2\over 2} \phi^2 +{\beta\over 3}
\phi^3+ {\lambda\over 4} \phi^4 + \sum_n \lambda_n
{\phi^{4+n}\over {M_p}^n}\, ,
\end{equation}
and then they assume that generically $\lambda_n = O(1)$, see e.g.
Eq. (128) in \cite{LythRiotto}.  If this assumption were correct,
one would have little control over the behavior of $V(\phi)$ at
$\phi > M_p$.

Here we have written $M_p$ explicitly, to expose the implicit
assumption made in \cite{LythRiotto}.  Why do we write $M_p$ in
the denominator, instead of $1000 M_p$? An intuitive reason is
that quantum gravity is non-renormalizable, so one should
introduce a cut-off at momenta $k \sim M_p$. This is a reasonable
assumption, but it does not imply validity of Eq.
(\ref{LythRiotto}). Indeed, the constant part of the scalar field
appears in the gravitational diagrams not directly, but only via
its effective potential $V(\phi)$ and the masses of particles
 interacting with the scalar field $\phi$. As a result,
the terms induced by quantum gravity effects are suppressed not by
factors ${\phi^n \over {M_p}^n}$, but by factors  $V\over {M_p}^4$
and $m^2(\phi)\over {M_p}^2$ \cite{book}. Consequently, quantum
gravity corrections to $V(\phi)$ become large not at $\phi
> M_p$, as one could infer from (\ref{LythRiotto}), but only at
super-Planckian energy density, or for super-Planckian masses.
This justifies our use of the simplest chaotic inflation models.

The simplest way to understand this argument is to consider the
case where the potential of the field $\phi$ is a constant,
$V=V_0$. Then the theory has a {\it shift symmetry}, $\phi \to
\phi +c$. This symmetry is not broken by perturbative quantum
gravity corrections, so no such terms as $\sum_n \lambda_n
{\phi^{4+n}\over {M_p}^n}$ are generated. This symmetry may be
broken by nonperturbative quantum gravity effects (wormholes?
virtual black holes?), but such effects, even if they exist, can
be made exponentially small \cite{Kallosh:1995hi}.

The idea of shift symmetry appears to be very fruitful in
application to inflation; we will return to it many times in this
paper. However, in some cases the scalar field $\phi$ itself may
have physical (geometric) meaning,  which may constrain the
possible values of the fields during inflation. The most important
example is given by $N = 1$ supergravity.

The effective potential of the complex scalar field $\Phi$ in
supergravity is given by the well-known  expression (in units $M_p
= 1$):
\begin{equation}\label{superpot}
V = e^{K} \left[K_{\Phi\bar\Phi}^{-1}\, |D_\Phi W|^2
-3|W|^2\right].
\end{equation}
Here $W(\Phi)$ is the superpotential, $\Phi$ denotes the scalar
component of the superfield  $\Phi$; $D_\Phi W= {\partial W\over
\partial \Phi} + {\partial K\over \partial \Phi} W$. The kinetic
term of the scalar field is given by $K_{\Phi\bar\Phi}\,
\partial_\mu \Phi \partial _\mu \bar\Phi$. The standard textbook
choice of the K\"ahler potential corresponding to the canonically
normalized fields $\Phi$ and $\bar\Phi$ is $K = \Phi\bar\Phi$, so
that $K_{\Phi\bar\Phi}=1$.

This immediately reveals a problem: At $\Phi > 1$ the potential is
extremely steep.  It blows up as $e^{|\Phi|^2}$, which makes it
very difficult to realize chaotic inflation in supergravity at
$\phi \equiv \sqrt 2|\Phi| > 1$. Moreover, the problem persists
even at small $\phi$. If, for example, one considers the simplest
case when there are many other scalar fields  in the theory  and
the superpotential does not depend on the inflaton field $\phi$,
then Eq. (\ref{superpot}) implies that at $\phi \ll 1$ the
effective mass of the inflaton field is $m^2_\phi = 3H^2$. This
violates the  condition $m^2_\phi \ll H^2$ required for successful
slow-roll inflation (so-called $\eta$-problem).

The major progress in SUGRA inflation during the last decade was
achieved in the context of the models of the hybrid inflation
type, where inflation may occur at $\phi \ll 1$. Among the best
models are the F-term inflation, where different contributions to
the effective mass term $m^2_\phi$ cancel \cite{F}, and D-term
inflation \cite{D}, where the dangerous term $e^K$ does not affect
the potential in the inflaton direction. A detailed discussion of
various versions of hybrid inflation in supersymmetric theories
can be found in \cite{LythRiotto}. A recent version of this
scenario, P-term inflation, which unifies F-term and D-term
models, was proposed in \cite{pterm}.

However, hybrid inflation occurs only on a relatively small energy
scale, and many of its versions do not lead to eternal inflation.
Therefore it would be nice to obtain inflation in a context of a
more general class of supergravity models.

This goal seemed very difficult to achieve; it took almost 20
years to find a natural realization of chaotic inflation model in
supergravity. Kawasaki, Yamaguchi and Yanagida suggested to take
the K\"ahler potential
\begin{equation} K = {1\over 2}(\Phi+\bar\Phi)^2
+X\bar X \end{equation}
 of the fields $\Phi$ and $X$, with the
superpotential $m\Phi X$ \cite{jap}.

At the first glance, this K\"ahler potential may seem somewhat
unusual. However, it can be obtained from the standard K\"ahler
potential $K =  \Phi \bar\Phi  +X\bar X$  by adding terms
$\Phi^2/2+\bar\Phi^2/2$, which do not give any contribution to the
kinetic term of the scalar fields $K_{\Phi\bar\Phi}\,
\partial_\mu \Phi \partial _\mu \bar\Phi$. In other words, the new
K\"ahler potential, just as the old one, leads to canonical
kinetic terms for the fields $\Phi$ and $X$, so it is as simple
and legitimate as the standard textbook K\"ahler potential.
However, instead of the U(1) symmetry with respect to rotation of
the field $\Phi$ in the complex plane, the new K\"ahler potential
has a {\it shift symmetry}; it does not depend on the imaginary
part of the field $\Phi$. The shift symmetry is broken only by the
superpotential.

This leads to a profound  change of the potential
(\ref{superpot}): the dangerous term $e^K$ continues growing
exponentially in the direction $(\Phi +\bar\Phi)$, but it remains
constant in the direction $(\Phi - \bar \Phi )$. Decomposing the
complex field $\Phi$ into two real scalar fields, $ \Phi = {1\over
\sqrt 2} (\eta +i\phi)$, one can find the resulting potential
$V(\phi,\eta,X)$ for $\eta, |X| \ll 1$:
\begin{equation}\label{superpot1}
V = {m^2\over 2} \phi^2 (1 + \eta^2) + m^2|X|^2.
\end{equation}
This potential has a deep valley, with a minimum at $\eta = X =0$.
Therefore the fields $\eta$ and $X$ rapidly fall down towards
$\eta = X =0$, after which the potential for the field $\phi$
becomes $V = {m^2\over 2} \phi^2$. This provides  a very simple
realization of eternal chaotic inflation scenario in supergravity
\cite{jap}. This model can be extended to include theories with
different power-law potentials, or models where inflation begins
as in the simplest versions of chaotic inflation scenario, but
ends as in new or hybrid inflation, see e.g.
\cite{Yamaguchi:2001pw,Yok}.

It is amazing that for almost 20 years nothing but inertia was
keeping us from using the version of the supergravity which was
free from the famous $\eta$ problem. As we will see shortly, the
situation with inflation in string theory is very similar, and may
have a similar resolution.

\section{Towards Inflation in String Theory}
\subsection{de Sitter vacua in string theory}

For a long time, it seemed rather difficult to obtain inflation in
M/string theory. The main problem here was the stability of
compactification of internal dimensions. For example, ignoring
non-perturbative effects to be discussed below, a typical
effective potential of the effective 4d theory obtained by
compactification in string theory of type IIB can be represented
in the following form:
\begin{equation}
V(\sigma,\rho,\phi) \sim e^{\sqrt 2\sigma -\sqrt6\rho}\ \tilde
V(\phi)
\end{equation}
Here $\sigma$ and $\rho$ are canonically normalized fields
representing the dilaton field and the volume of the compactified
space; $\phi$ stays for all other fields.

If $\sigma$ and $\rho$ were constant, then the potential $\tilde
V(\phi)$ could drive inflation.  However, this does not happen
because of the steep exponent $e^{\sqrt 2\sigma -\sqrt6\rho}$,
which rapidly pushes the dilaton field $\sigma$ to $-\infty$, and
the volume modulus $\rho$ to $+\infty$. As a result, the radius of
compactification becomes infinite; instead of inflating, 4d space
decompactifies and becomes 10d.

Thus in order to describe inflation one should first learn how to
stabilize the dilaton and the volume modulus. The dilaton
stabilization was achieved in \cite{GKP}. The most difficult
problem was to stabilize the volume. The solution of this problem
was found in \cite{KKLT} (KKLT construction). It consists of two
steps.

First of all, due to a combination of effects related to warped
geometry of the compactified space and nonperturbative effects
calculated directly in 4d (instead of being obtained by
compactification), it was possible to obtain a supersymmetric AdS
minimum of the effective potential for $\rho$. This fixed the
volume modulus, but in a state with a negative vacuum energy. Then
we added an anti-${D3}$ brane with the positive energy $\sim
\rho^{-3}$. This addition uplifted the minimum of the potential to
the state with a positive vacuum energy.

Instead of adding an anti-${D3}$ brane, which explicitly breaks
supersymmetry, one can add  a D7 brane with fluxes. This results
in the appearance of a D-term which has a similar dependence on
$\rho$, but leads to spontaneous supersymmetry breaking
\cite{Burgess:2003ic}. In either case, one ends up with a
metastable dS state which can decay by tunnelling and formation of
bubbles of 10d space with vanishing vacuum energy density. The
decay rate is extremely small \cite{KKLT}, so for all practical
purposes, one obtains an exponentially expanding de Sitter space
with the stabilized volume of the internal space.\footnote{It is
also possible to find de Sitter solutions in noncritical string
theory \cite{Str}.}

\subsection{Inflation in string theory and shift symmetry}

During the last few years there were many suggestions how to
obtain hybrid inflation in string theory by considering motion of
branes in the compactified space, see \cite{Dvali:1998pa,Quevedo}
and references therein. The main problem of all of these models
was the absence of stabilization of the compactified space. Once
this problem was solved for dS space \cite{KKLT}, one could try to
revisit these models and develop models of brane inflation
compatible with the volume stabilization.

The first idea \cite{KKLMMT} was to consider a pair of D3 and
anti-D3 branes in the warped geometry studied in \cite{KKLT}. The
role of the inflaton field could be played by the interbrane
separation. A description of this situation in terms of the
effective 4d supergravity involved K\"ahler potential
\begin{equation}
K = -3\log (\rho+\bar\rho -k(\phi,\bar\phi)), \end{equation}
 where the function
$k(\phi,\bar\phi)$ for the inflaton field $\phi$, at small $\phi$,
was taken in the simplest form $k(\phi,\bar\phi)= \phi\bar\phi$.
 If one makes  the simplest
assumption that the superpotential does not depend on $\phi$, then
the $\phi$ dependence of the potential (\ref{superpot})  comes
from the term $e^K =(\rho+\bar\rho - \phi\bar\phi)^{-3}$.
Expanding this term near the  stabilization point $\rho = \rho_0$,
one finds that the inflaton field has a mass $m^2_\phi = 2H^2$.
Just like the similar relation $m^2_\phi = 3H^2$ in the simplest
models of supergravity, this is not what we want for inflation.

One way to solve this problem is to consider $\phi$-dependent
superpotentials. By doing so, one may fine-tune $m^2_\phi$ to be
$O(10^{-2}) H^2$ in a vicinity of the point where inflation occurs
\cite{KKLMMT}. Whereas fine-tuning is certainly undesirable, in
the context of string cosmology it may not be a serious drawback.
Indeed, if there exist many realizations of string theory
\cite{Douglas}, then one might argue that all realizations not
leading to inflation can be discarded, because they do not
describe a universe in which we could live. Meanwhile, those
non-generic realizations, which lead to eternal inflation,
describe inflationary universes with an indefinitely large and
ever-growing volume of inflationary domains. This makes the issue
of fine-tuning less problematic.\footnote{One should note that
decreasing of $m_\phi^2$ is not the only way to get inflation; one
may reach the same goal by considering theories with non-minimal
kinetic terms, see e.g. \cite{kinfl,Dim,Silverstein:2003hf}.}

Can we avoid fine-tuning altogether? One of the possible ideas is
to find theories with some kind of shift symmetry. Another
possibility is to construct something like D-term inflation, where
the flatness of the potential is not spoiled by the term $e^K$.
Both of these ideas were explored in a recent paper
\cite{Hsu:2003cy} based on the model of D3/D7 inflation in string
theory \cite{renata}. In this model the K\"ahler potential is
given by
\begin{equation}
K = -3\log (\rho+\bar\rho) -{1\over 2}(\phi - \bar\phi)^2,
\end{equation}
and superpotential depends only on $\rho$. The shift symmetry
$\phi \to \phi+c$ in this model is related to the requirement of
unbroken supersymmetry of branes in a BPS state.

The effective potential with respect to the field $\rho$ in this
model coincides with the KKLT potential
\cite{KKLT,Burgess:2003ic}. In the direction of the real part of
the field $\phi$, which can be considered an inflaton, the
potential is exactly flat, until one adds other fields which break
this flatness due to quantum corrections and produce a potential
similar to the potential of D-term inflation \cite{Hsu:2003cy}.
The origin of the shift symmetry of the K\"ahler potential in
certain string models was revealed in a recent paper
\cite{Angelantonj:2003up}. It is related to special geometry of
extended supergravity. This shift symmetry was also studied in a
recent version of D3/D7 inflation proposed in
\cite{Koyama:2003yc}.

Shift symmetry may help to obtain inflation in other models as
well. For example, one may explore the possibility of using the
K\"ahler potential $K = -3\log (\rho+\bar\rho -{1\over 2}(\phi -
\bar\phi)^2))$ instead of the potential used in \cite{KKLMMT}. The
modified K\"ahler potential does not depend on the real part of
the field $\phi$, which can be considered an inflaton. Therefore
the dangerous term $m^2_\phi = 2H^2$ vanishes, i.e. the main
obstacle to the consistent brane inflation in the model of Ref.
\cite{KKLMMT} disappears! A discussion of the possibility to
implement shift symmetry in the model of Ref. \cite{KKLMMT} can be
also found in \cite{Tye}.

However, for a while it still remained unclear whether shift
symmetry is just a condition which we want to impose on the theory
in order to get inflation, or an unavoidable property of the
theory, which remains valid even after the KKLT volume
stabilization. The answer to this question was found only very
recently, and it appears to be model-dependent. It was shown in
\cite{kalrecent} that in a certain class of models, including
D3/D7 models
\cite{renata,Hsu:2003cy,Angelantonj:2003up,Koyama:2003yc}, the
shift symmetry of the effective 4d theory is not an assumption but
an unambiguous consequence of the underlying mathematical
structure of the theory. This may allow us to obtain a natural
realization of inflation in string theory.

\section{Eternal inflation and stringy landscape}\label{land}

Even though we are still at the very first stages of implementing
inflation in string theory, it is very tempting to speculate about
possible generic features and consequences of such a construction.

First of all, KKLT construction shows that the vacuum energy after
the volume stabilization is a function of many different
parameters in the theory. One may wonder how many different
choices do we actually have. There were many attempts to
investigate this issue. For example, many years ago it was argued
\cite{Duff} that there are infinitely many $AdS_4 \times X7$ vacua
of D=11 supergravity. An early estimate of the total number of
different 4d string vacua gave the number $10^{1500}$
\cite{Lerche}. At present we are more interested in counting
different flux vacua \cite{BP,Douglas}, where the possible
numbers, depending on specific assumptions, may vary in the range
from $10^{20}$ to $10^{1000}$. Some of these vacuum states with
positive vacuum energy can be stabilized using the KKLT approach.
Each of such states will correspond to a metastable vacuum state.
It decays within a cosmologically large time, which is, however,
smaller than  the `recurrence time' $e^{S(\phi)} $, where $S(\phi)
= {24\pi^2\over V(\phi)}$ is the entropy of dS space with the
vacuum energy density $V(\phi)$ \cite{KKLT}.

But this is not the whole story; old inflation does not describe
our world. In addition to these metastable vacuum states, there
should exist various slow-roll inflationary solutions, where the
properties of the system practically do not change during the
cosmological time $H^{-1}$. It might happen that such states,
corresponding to flat directions in the string theory landscape,
exist not only during inflation in the very early universe, but
also at the present stage of the accelerated expansion of the
universe. This would simplify obtaining an anthropic solution of
the cosmological constant problem along the lines of
\cite{Linde84,BP}.

If the slow-roll condition $V'' \ll V$ is satisfied all the way
from one dS minimum of the effective potential to another, then
one can show, using stochastic approach to inflation, that the
probability to find the field $\phi$ at any of these minima, or at
any given point between them,  is proportional to $e^{S(\phi)}$.
In other words, the relative probability to find the field taking
some value $\phi_1$ as compared to some other value $\phi_0$, is
proportional to $e^{\Delta S} = e^{S(\phi_1)-S(\phi_0)}$
\cite{Open,KKLT}. One may argue, using Euclidean approach, that
this simple thermodynamic relation should remain valid for the
relative probability to find a given point in any of the
metastable dS vacua, even if the trajectory between them does not
satisfy the slow-roll condition $m^2 \ll H^2$
\cite{HM,Lee:qc,Garriga:1997ef,Dyson:2002pf}.

The resulting picture resembles eternal inflation in the old
inflation scenario. However, now we have an incredibly large
number of false vacuum states, plus some states which may allow
slow-roll inflation. Once inflation begins, different parts of the
universe start wondering from one of these vacuum states to
another, so that the universe becomes divided into indefinitely
many regions with all possible laws of low-energy physics
corresponding to different 4d vacua of string theory \cite{book}.

As we already argued, the best inflationary scenario would
describe a slow-roll eternal inflation starting at the maximal
possible energy density (minimal dS entropy). It would be almost
as good to have a low-energy slow-roll eternal inflation. Under
certain conditions, such regimes may exist in string theory
\cite{KKLMMT}. However, whereas any of these regimes would make us
happy, we already have something that can make us smile.
Multi-level eternal inflation of the old inflation type, which
appears in string theory in the context of the KKLT construction,
may be very useful being combined with the slow-roll inflation,
even if the slow-roll inflation by itself is not eternal. We will
give a particular example, which is very similar to the one
considered in \cite{Linde:1987yb}.

Suppose we have two noninteracting scalar fields: field $\phi$
with the potential of the old inflation type, and field $\chi$
with the potential which may lead to a slow-roll inflation. Let us
assume that the slow-roll inflation belongs to the worst case
scenario discussed in Section \ref{alt}: it occurs only on low
energy scale, and it is not eternal. How can we provide initial
conditions for such a low-scale inflation?

Let us assume that the Hubble constant at the stage of old
inflation is much greater than the curvature of the potential
which drives the slow-roll inflation. (This is a natural
assumption, considering huge number of possible dS states, and the
presumed smallness of energy scale of the slow-roll inflation.) In
this case large inflationary fluctuations of the field $\chi$ will
be generated during eternal old inflation. These fluctuations will
give the field $\chi$ different values in different exponentially
large parts of the universe. When old inflation ends, there will
be many practically homogeneous parts of the universe where  the
field $\chi$ will take  values corresponding to good initial
conditions for a slow-roll inflation. Then the relative fraction
of the volume of such parts will grow exponentially.

Moreover, as it was argued in \cite{LLM}, the probability (per
unit time and unit volume) to jump back to the eternally inflating
regime is always ~{\it finite}, even after the field enters the
regime where, naively, one would not expect eternal inflation.
Each bubble of a new phase which appears during the decay of the
eternally inflating dS space is an open universe of an {\it
infinite} volume. Therefore during the slow-roll inflation there
always will be some inflationary domains  jumping back to the
original dS space, so some kind of stationary equilibrium will
always exist between various parts of the inflationary universe.

Thus, the existence of many different dS vacua in string theory
leads to the regime of eternal inflation. This regime may help us
to solve the problem of initial conditions for the slow-roll
inflation even in the models where the slow-roll inflation by
itself is not eternal and would occur only on a small energy
scale.

I  am very grateful to the organizers of the Nobel Symposium
``Cosmology \& String Theory" for the hospitality. It is a special
pleasure to express my gratitude to T. Banks, S. Kachru, R.
Kallosh, L. Kofman, V. Mukhanov, E. Silverstein, and  L. Susskind
for many enlightening discussions. This work was supported by NSF
grant PHY-0244728.



\begin{thebibliography}{99}


\bibitem{Star} A.~A.~Starobinsky,
``Spectrum Of Relict Gravitational Radiation And The Early State
Of The  Universe,'' JETP Lett.\  {\bf 30}, 682 (1979) [Pisma Zh.\
Eksp.\ Teor.\ Fiz.\  {\bf 30}, 719 (1979)]; A.~A.~Starobinsky, ``A
New Type Of Isotropic Cosmological Models Without Singularity,''
Phys.\ Lett.\ B {\bf 91}, 99 (1980).


\bibitem{Guth} A.~H.~Guth,
``The Inflationary Universe: A Possible Solution To The Horizon
And Flatness Problems,'' Phys.\ Rev.\ D {\bf 23}, 347 (1981).


\bibitem{Kirzhnits}
 D.~A.~Kirzhnits and A.~D.~Linde,
``Symmetry Behavior In Gauge Theories,'' Annals Phys.\  {\bf 101},
195 (1976).



\bibitem{book} A.D. Linde,  {\it  Particle  Physics  and
Inflationary Cosmology} (Harwood, Chur, Switzerland, 1990).




\bibitem{New}
A.~D.~Linde, ``A New Inflationary Universe Scenario: A Possible
Solution Of The Horizon, Flatness, Homogeneity, Isotropy And
Primordial Monopole Problems,'' Phys.\ Lett.\ B {\bf 108}, 389
(1982); A.~Albrecht and P.~J.~Steinhardt, ``Cosmology For Grand
Unified Theories With Radiatively Induced Symmetry Breaking,''
Phys.\ Rev.\ Lett.\  {\bf 48}, 1220 (1982).




\bibitem{Mukh} V.~F.~Mukhanov and G.~V.~Chibisov,
``Quantum Fluctuation And `Nonsingular' Universe,'' JETP Lett.\
{\bf 33}, 532 (1981) [Pisma Zh.\ Eksp.\ Teor.\ Fiz.\  {\bf 33},
549 (1981)].




\bibitem{Hawk} S.~W.~Hawking,
``The Development Of Irregularities In A Single Bubble
Inflationary Universe,'' Phys.\ Lett.\ B {\bf 115}, 295 (1982);
A.~A.~Starobinsky, ``Dynamics Of Phase Transition In The New
Inflationary Universe Scenario And Generation Of Perturbations,''
Phys.\ Lett.\ B {\bf 117}, 175 (1982); A.~H.~Guth and S.~Y.~Pi,
``Fluctuations In The New Inflationary Universe,'' Phys.\ Rev.\
Lett.\  {\bf 49}, 1110 (1982); J.~M.~Bardeen, P.~J.~Steinhardt and
M.~S.~Turner, ``Spontaneous Creation Of Almost Scale - Free
Density Perturbations In An Inflationary Universe,'' Phys.\ Rev.\
D {\bf 28}, 679 (1983).

\bibitem{Mukh2} V.~F.~Mukhanov,
``Gravitational Instability Of The Universe Filled With A Scalar
Field,'' JETP Lett.\  {\bf 41}, 493 (1985) [Pisma Zh.\ Eksp.\
Teor.\ Fiz.\  {\bf 41}, 402 (1985)]; V.~F.~Mukhanov, H.~A.~Feldman
and R.~H.~Brandenberger, ``Theory Of Cosmological Perturbations,''
Phys.\ Rept.\  {\bf 215}, 203 (1992).



\bibitem{Chaot} A.~D.~Linde,
``Chaotic Inflation,'' Phys.\ Lett.\ B {\bf 129}, 177 (1983).



\bibitem{DL}
A.~D.~Dolgov and A.~D.~Linde, ``Baryon Asymmetry In Inflationary
Universe,'' Phys.\ Lett.\ B {\bf 116}, 329 (1982); L.~F.~Abbott,
E.~Farhi and M.~B.~Wise, ``Particle Production In The New
Inflationary Cosmology,'' Phys.\ Lett.\ B {\bf 117}, 29 (1982).

\bibitem{KLS} L.~Kofman, A.~D.~Linde and A.~A.~Starobinsky,
``Reheating after inflation,'' Phys.\ Rev.\ Lett.\  {\bf 73}, 3195
(1994) [arXiv:hep-th/9405187]; L.~Kofman, A.~D.~Linde and
A.~A.~Starobinsky, ``Towards the theory of reheating after
inflation,'' Phys.\ Rev.\ D {\bf 56}, 3258 (1997)
[arXiv:hep-ph/9704452].


\bibitem{tach}  G.~N.~Felder, J.~Garcia-Bellido, P.~B.~Greene,
L.~Kofman, A.~D.~Linde and I.~Tkachev, ``Dynamics of symmetry
breaking and tachyonic preheating,'' Phys.\ Rev.\ Lett.\  {\bf
87}, 011601 (2001) [arXiv:hep-ph/0012142]; G.~N.~Felder, L.~Kofman
and A.~D.~Linde, ``Tachyonic instability and dynamics of
spontaneous symmetry breaking,'' Phys.\ Rev.\ D {\bf 64}, 123517
(2001) [arXiv:hep-th/0106179].



\bibitem{OvrStein}
B.~A.~Ovrut and P.~J.~Steinhardt, ``Supersymmetry And Inflation: A
New Approach,'' Phys.\ Lett.\ B {\bf 133}, 161 (1983); B.~A.~Ovrut
and P.~J.~Steinhardt, ``Inflationary Cosmology And The Mass
Hierarchy In Locally Supersymmetric Theories,'' Phys.\ Rev.\
Lett.\  {\bf 53}, 732 (1984); B.~A.~Ovrut and P.~J.~Steinhardt,
``Locally Supersymmetric Cosmology And The Gauge Hierarchy,''
Phys.\ Rev.\ D {\bf 30}, 2061 (1984); B.~A.~Ovrut and
P.~J.~Steinhardt, ``Supersymmetric Inflation, Baryon Asymmetry And
The Gravitino Problem,'' Phys.\ Lett.\ B {\bf 147}, 263 (1984).

\bibitem{Linde:cd}
A.~D.~Linde, ``Primordial Inflation Without Primordial
Monopoles,'' Phys.\ Lett.\ B {\bf 132} (1983) 317.

\bibitem{Hybrid}
A.~D.~Linde, ``Axions in inflationary cosmology,'' Phys.\ Lett.\ B
{\bf 259}, 38 (1991); A.~D.~Linde, ``Hybrid inflation,'' Phys.\
Rev.\ D {\bf 49}, 748 (1994) [astro-ph/9307002].




\bibitem{Vilenkin:wt}
A.~Vilenkin and L.~H.~Ford, ``Gravitational Effects Upon
Cosmological Phase Transitions,'' Phys.\ Rev.\ D {\bf 26}, 1231
(1982).


\bibitem{Linde:uu}
A.~D.~Linde, ``Scalar Field Fluctuations In Expanding Universe And
The New Inflationary Universe Scenario,'' Phys.\ Lett.\ B {\bf
116}, 335 (1982).


\bibitem{Vilenkin:xq}
P.~J.~Steinhardt, ``Natural Inflation,'' In: {\it  The Very Early
Universe}, ed. G.W. Gibbons, S.W. Hawking and S.Siklos, Cambridge
University Press, (1983); A.~D.~Linde, ``Nonsingular Regenerating
Inflationary Universe,''  Cambridge University preprint
Print-82-0554 (1982); A.~Vilenkin, ``The Birth Of Inflationary
Universes,'' Phys.\ Rev.\ D {\bf 27}, 2848 (1983).




\bibitem{Eternal}
A.~D.~Linde, ``Eternally Existing Selfreproducing Chaotic
Inflationary Universe,'' Phys.\ Lett.\ B {\bf 175}, 395 (1986).

\bibitem{LLM}
A.~D.~Linde, D.~A.~Linde and A.~Mezhlumian, ``From the Big Bang
theory to the theory of a stationary universe,'' Phys.\ Rev.\ D
{\bf 49}, 1783 (1994) [arXiv:gr-qc/9306035].


\bibitem{landscape}
L.~Susskind, ``The anthropic landscape of string theory,''
arXiv:hep-th/0302219.

\bibitem{Douglas}
M.~R.~Douglas, ``The statistics of string / M theory vacua,'' JHEP
{\bf 0305}, 046 (2003) [arXiv:hep-th/0303194]; S.~Ashok and
M.~R.~Douglas, ``Counting flux vacua,'' arXiv:hep-th/0307049;
M.~R.~Douglas, ``Statistics of String vacua,''
arXiv:hep-ph/0401004.


\bibitem{Dodelson:2003ip}
S.~Dodelson, ``Coherent phase argument for inflation,'' AIP Conf.\
Proc.\  {\bf 689}, 184 (2003) [arXiv:hep-ph/0309057].

\bibitem{Mukhanov:2003xr}
V.~Mukhanov, ``CMB-slow, or How to Estimate Cosmological
Parameters by Hand,'' arXiv:astro-ph/0303072.

\bibitem{WMAP} H.~V.~Peiris {\it et al.},
``First year Wilkinson Microwave Anisotropy Probe (WMAP)
observations: Implications for inflation,'' Astrophys.\ J.\
Suppl.\  {\bf 148}, 213 (2003) [arXiv:astro-ph/0302225].

\bibitem{Tegmark}
M.~Tegmark {\it et al.}  [SDSS Collaboration], ``Cosmological
parameters from SDSS and WMAP,'' arXiv:astro-ph/0310723.

\bibitem{Efstathiou:2003wr}
G.~Efstathiou, ``The Statistical Significance of the Low CMB
Mulitipoles,'' arXiv:astro-ph/0306431; G.~Efstathiou, ``A Maximum
Likelihood Analysis of the Low CMB Multipoles from WMAP,''
arXiv:astro-ph/0310207.



\bibitem{Contaldi}
C.~R.~Contaldi, M.~Peloso, L.~Kofman and A.~Linde, ``Suppressing
the lower Multipoles in the CMB Anisotropies,'' JCAP {\bf 0307},
002 (2003) [arXiv:astro-ph/0303636].




\bibitem{SperTur} D. Spergel and N. Turok, ``Textures and cosmic structure,'' Scientific
American {\bf 266}, 52 (1992).



\bibitem{KOST}
J.~Khoury, B.~A.~Ovrut, P.~J.~Steinhardt and N.~Turok, ``The
ekpyrotic universe: Colliding branes and the origin of the hot big
bang,'' Phys.\ Rev.\ D {\bf 64}, 123522 (2001)
[arXiv:hep-th/0103239].



\bibitem{KKL}
R.~Kallosh, L.~Kofman and A.~D.~Linde, ``Pyrotechnic universe,''
Phys.\ Rev.\ D {\bf 64}, 123523 (2001) [arXiv:hep-th/0104073].



\bibitem{KKLTS}
R.~Kallosh, L.~Kofman, A.~D.~Linde and A.~A.~Tseytlin, ``BPS
branes in cosmology,'' Phys.\ Rev.\ D {\bf 64}, 123524 (2001)
[arXiv:hep-th/0106241].



\bibitem{cyclic}
P.~J.~Steinhardt and N.~Turok, ``Cosmic evolution in a cyclic
universe,'' Phys.\ Rev.\ D {\bf 65}, 126003 (2002)
[arXiv:hep-th/0111098].

\bibitem{Liu:2002ft}
H.~Liu, G.~Moore and N.~Seiberg, ``Strings in a time-dependent
orbifold,'' JHEP {\bf 0206}, 045 (2002) [arXiv:hep-th/0204168].

\bibitem{Horowitz:2002mw}
G.~T.~Horowitz and J.~Polchinski, ``Instability of spacelike and
null orbifold singularities,'' Phys.\ Rev.\ D {\bf 66}, 103512
(2002) [arXiv:hep-th/0206228].


\bibitem{extended}
D.~La and P.~J.~Steinhardt, ``Extended Inflationary Cosmology,''
Phys.\ Rev.\ Lett.\  {\bf 62}, 376 (1989) [Erratum-ibid.\  {\bf
62}, 1066 (1989)].

\bibitem{natural}
K.~Freese, J.~A.~Frieman and A.~V.~Olinto,
 ``Natural Inflation With Pseudo - Nambu-Goldstone Bosons,''
Phys.\ Rev.\ Lett.\  {\bf 65}, 3233 (1990).

\bibitem{Lyth:2003kp}
D.~H.~Lyth, ``Which is the best inflation model?,''
arXiv:hep-th/0311040.


\bibitem{Kallosh:1995hi}
R.~Kallosh, A.~Linde, D.~Linde and L.~Susskind, ``Gravity and
global symmetries,'' Phys.\ Rev.\  {\bf D52}, 912 (1995)
[hep-th/9502069].


\bibitem{F}
E.~J.~Copeland, A.~R.~Liddle, D.~H.~Lyth, E.~D.~Stewart and
D.~Wands, ``False vacuum inflation with Einstein gravity,'' Phys.\
Rev.\ D {\bf 49}, 6410 (1994) [astro-ph/9401011]; G.~R.~Dvali,
Q.~Shafi and R.~Schaefer, ``Large scale structure and
supersymmetric inflation without fine tuning,'' Phys.\ Rev.\
Lett.\  {\bf 73}, 1886 (1994) [hep-ph/9406319];
 A.~D.~Linde and A.~Riotto,
``Hybrid inflation in supergravity,'' Phys.\ Rev.\ D {\bf 56},
1841 (1997) [arXiv:hep-ph/9703209].


\bibitem{D}
P.~Binetruy and G.~Dvali, ``D-term inflation,'' Phys.\ Lett.\ B
{\bf 388}, 241 (1996) [hep-ph/9606342]; E.~Halyo, ``Hybrid
inflation from supergravity D-terms,'' Phys.\ Lett.\ B {\bf 387},
43 (1996) [hep-ph/9606423].


\bibitem{LythRiotto}
D.~H.~Lyth and A.~Riotto, ``Particle physics models of inflation
and the cosmological density  perturbation,'' Phys.\ Rept.\  {\bf
314}, 1 (1999) [hep-ph/9807278].



\bibitem{pterm}
R.~Kallosh and A.~Linde, ``P-term, D-term and F-term inflation,''
JCAP {\bf 0310}, 008 (2003) [arXiv:hep-th/0306058].



\bibitem{jap}
M.~Kawasaki, M.~Yamaguchi and T.~Yanagida, ``Natural chaotic
inflation in supergravity,'' Phys.\ Rev.\ Lett.\  {\bf 85}, 3572
(2000) [arXiv:hep-ph/0004243].

\bibitem{Yamaguchi:2001pw}
M.~Yamaguchi and J.~Yokoyama, ``New inflation in supergravity with
a chaotic initial condition,'' Phys.\ Rev.\ D {\bf 63}, 043506
(2001) [arXiv:hep-ph/0007021]; M.~Yamaguchi, ``Natural double
inflation in supergravity,'' Phys.\ Rev.\ D {\bf 64}, 063502
(2001) [arXiv:hep-ph/0103045].


\bibitem{Yok}
M.~Yamaguchi and J.~Yokoyama, ``Chaotic hybrid new inflation in
supergravity with a running spectral
arXiv:hep-ph/0307373.


\bibitem{GKP}
S.~B.~Giddings, S.~Kachru and J.~Polchinski, ``Hierarchies from
fluxes in string compactifications,'' Phys. Rev. {\bf D66}, 106006
(2002) [arXiv:hep-th/0105097].


\bibitem{KKLT}
S.~Kachru, R.~Kallosh, A.~Linde and S.~P.~Trivedi, ``De Sitter
vacua in string theory,'' Phys.\ Rev.\ D {\bf 68}, 046005 (2003)
[arXiv:hep-th/0301240].


\bibitem{Burgess:2003ic}
C.~P.~Burgess, R.~Kallosh and F.~Quevedo, ``de Sitter string vacua
from supersymmetric D-terms,'' JHEP {\bf 0310}, 056 (2003)
[arXiv:hep-th/0309187].

\bibitem{Str} E.~Silverstein,
``(A)dS backgrounds from asymmetric orientifolds,''
arXiv:hep-th/0106209;
A.~Maloney, E.~Silverstein and A.~Strominger, ``De Sitter space in
noncritical string theory,'' arXiv:hep-th/0205316.

\bibitem{Dvali:1998pa}
G.~R.~Dvali and S.~H.~H.~Tye, ``Brane inflation,'' Phys.\ Lett.\ B
{\bf 450}, 72 (1999) [arXiv:hep-ph/9812483].


\bibitem{Quevedo}
F. Quevedo, ``Lectures on String/Brane Cosmology,''
hep-th/0210292.



\bibitem{KKLMMT}
S.~Kachru, R.~Kallosh, A.~Linde, J.~Maldacena, L.~McAllister and
S.~P.~Trivedi, ``Towards inflation in string theory,'' JCAP {\bf
0310}, 013 (2003) [arXiv:hep-th/0308055].

\bibitem{kinfl}
C.~Armendariz-Picon, T.~Damour and V.~Mukhanov, ``k-inflation,''
Phys.\ Lett.\ B {\bf 458}, 209 (1999) [arXiv:hep-th/9904075].


\bibitem{Dim}
S.~Dimopoulos and S.~Thomas, ``Discretuum versus continuum dark
energy,'' Phys.\ Lett.\ B {\bf 573}, 13 (2003)
[arXiv:hep-th/0307004].

\bibitem{Silverstein:2003hf}
E.~Silverstein and D.~Tong, ``Scalar speed limits and cosmology:
Acceleration from D-cceleration,'' arXiv:hep-th/0310221.

\bibitem{Hsu:2003cy}
J.~P.~Hsu, R.~Kallosh and S.~Prokushkin, ``On brane inflation with
volume stabilization,'' JCAP {\bf 0312}, 009 (2003)
[arXiv:hep-th/0311077].



\bibitem{renata}
R.~Kallosh, ``N = 2 supersymmetry and de Sitter space,''
arXiv:hep-th/0109168; C.~Herdeiro, S.~Hirano and R.~Kallosh,
``String theory and hybrid inflation/acceleration,'' JHEP {\bf
0112} (2001) 027 [arXiv:hep-th/0110271]; K.~Dasgupta, C.~Herdeiro,
S.~Hirano and R.~Kallosh, ``D3/D7 inflationary model and
M-theory,'' Phys.\ Rev.\ D {\bf 65}, 126002 (2002)
[arXiv:hep-th/0203019].


\bibitem{Angelantonj:2003up}
C.~Angelantonj, R.~D'Auria, S.~Ferrara and M.~Trigiante, ``K3 x
T**2/Z(2) orientifolds with fluxes, open string moduli and
critical points,'' arXiv:hep-th/0312019.

\bibitem{Koyama:2003yc}
F.~Koyama, Y.~Tachikawa and T.~Watari, ``Supergravity analysis of
hybrid inflation model from D3-D7 system,'' arXiv:hep-th/0311191.

\bibitem{Tye}
H.~Firouzjahi and S.~H.~H.~Tye, ``Closer towards inflation in
string theory,'' arXiv:hep-th/0312020.


\bibitem{kalrecent} J. P. Hsu and R. Kallosh, ``Volume Stabilization and the Origin of the Inflaton Shift Symmetry in
  String Theory,'' hep-th/0402047.



\bibitem{Duff}
M.~J.~Duff, B.~E.~W.~Nilsson and C.~N.~Pope, ``Kaluza-Klein
Supergravity,'' Phys.\ Rept.\  {\bf 130}, 1 (1986).


\bibitem{Lerche}
W.~Lerche, D.~Lust and A.~N.~Schellekens, ``Chiral
Four-Dimensional Heterotic Strings From Selfdual Lattices,''
Nucl.\ Phys.\ B {\bf 287}, 477 (1987); M.~J.~Duff, ``Not The
Standard Superstring Review,'' CERN-TH-4749-87 {\it Proc. of Int.
School of Subnuclear Physics, The Super-World II, Erice, Italy,
Aug 6-14, 1987}.

\bibitem{BP}
R.~Bousso and J.~Polchinski, ``Quantization of four-form fluxes
and dynamical neutralization of the  cosmological constant,'' JHEP
{\bf 0006}, 006 (2000) [arXiv:hep-th/0004134].

\bibitem{Open}
A.~D.~Linde, ``Quantum creation of an open inflationary
universe,'' Phys.\ Rev.\ D {\bf 58}, 083514 (1998)
[arXiv:gr-qc/9802038].



\bibitem{Linde84}
A.~D.~Linde, ``The Inflationary Universe,'' Rept.\ Prog.\ Phys.\
{\bf 47}, 925 (1984); A.~D.~Sakharov, ``Cosmological Transitions
With A Change In Metric Signature,'' Sov.\ Phys.\ JETP {\bf 60},
214 (1984) [Zh.\ Eksp.\ Teor.\ Fiz.\  {\bf 87}, 375 (1984)]; A.D.
Linde, ``Inflation And Quantum Cosmology,'' Print-86-0888 (June
1986) in {\it 300 Years of Gravitation}, ed. by S.W. Hawking and
W. Israel, Cambridge University Press, Cambridge (1987);
S.~Weinberg, ``Anthropic Bound On The Cosmological Constant,''
Phys.\ Rev.\ Lett.\  {\bf 59}, 2607 (1987); A.~Vilenkin,
``Predictions From Quantum Cosmology,'' Phys.\ Rev.\ Lett.\  {\bf
74}, 846 (1995) [arXiv:gr-qc/9406010]; G. Efstathiou, MNRAS {\bf
274}, L73 (1995); H.~Martel, P.~R.~Shapiro and S.~Weinberg,
``Likely Values of the Cosmological Constant,'' Astrophys.\ J.\
{\bf 492}, 29 (1998) [arXiv:astro-ph/9701099]; J.~Garriga and
A.~Vilenkin, ``On likely values of the cosmological constant,''
Phys.\ Rev.\ D {\bf 61}, 083502 (2000) [arXiv:astro-ph/9908115];
J.~L.~Feng, J.~March-Russell, S.~Sethi and F.~Wilczek, ``Saltatory
relaxation of the cosmological constant,'' Nucl.\ Phys.\ B {\bf
602}, 307 (2001) [arXiv:hep-th/0005276].

\bibitem{HM}
S.~W.~Hawking and I.~G.~Moss, ``Supercooled Phase Transitions In
The Very Early Universe,'' Phys.\ Lett.\ B {\bf 110}, 35 (1982).

\bibitem{Lee:qc}
K.~M.~Lee and E.~J.~Weinberg, ``Decay Of The True Vacuum In Curved
Space-Time,'' Phys.\ Rev.\ D {\bf 36}, 1088 (1987).

\bibitem{Garriga:1997ef}
J.~Garriga and A.~Vilenkin, ``Recycling universe,'' Phys.\ Rev.\ D
{\bf 57}, 2230 (1998) [arXiv:astro-ph/9707292].

\bibitem{Dyson:2002pf}
L.~Dyson, M.~Kleban and L.~Susskind, ``Disturbing implications of
a cosmological constant,'' JHEP {\bf 0210}, 011 (2002)
[arXiv:hep-th/0208013]; N.~Goheer, M.~Kleban and L.~Susskind,
``The trouble with de Sitter space,'' JHEP {\bf 0307}, 056 (2003)
[arXiv:hep-th/0212209].

\bibitem{Linde:1987yb}
A.~D.~Linde, ``Chaotic Inflation With Constrained Fields,'' Phys.\
Lett.\ B {\bf 202}, 194 (1988).




\end{thebibliography}
\end{document}